\documentclass{article}
\usepackage{spconf,amsmath,amssymb,graphicx,booktabs}
\usepackage[hidelinks]{hyperref}  
\usepackage{xcolor}

\usepackage[utf8]{inputenc}
\usepackage[T1]{fontenc}
\usepackage{graphicx}
\usepackage{float}   
\usepackage{booktabs}
\usepackage{multirow}
\usepackage{amsmath}
\usepackage{xcolor}
\usepackage[hidelinks]{hyperref}

\newcommand{\dgrow}{\ensuremath{\Delta_{\mathrm{grow}}}}

\definecolor{figgray}{HTML}{546E7A}

\makeatletter  
\renewcommand\section{\@startsection{section}{1}{\z@}%
  {-2.5ex plus -1ex minus -.2ex}{1.5ex plus .2ex}{}}
\renewcommand\subsection{\@startsection{subsection}{2}{\z@}%
  {-2.25ex plus -1ex minus -.2ex}{1ex plus .2ex}{}}
\makeatother

\newif\ifdraftred
\draftredfalse
\ifdraftred
  \AtBeginDocument{\color{red}}
\fi
\graphicspath{{figures/}}

\title{Neuron-Level Architecture Growth:\\
A Controlled Evaluation for EEG Time-Series Decoding}

\name{%
  Adam Mounir$^{1,2}$ \hspace{0.3em}
  Stella Douka$^{1}$ \hspace{0.3em}
  Arnault H. Caillet$^{2,4}$ \hspace{0.3em}
  Bruno Aristimunha$^{2,3,*}$ \hspace{0.3em}
  Sylvain Chevallier$^{1,*}$
  \thanks{$^{*}$Equal supervision.}%
}
\address{%
  $^{1}$Inria TAU -- LISN, Universit\'e Paris-Saclay, France \quad
  $^{2}$Yneuro, Paris, France \\
  $^{3}$University of California San Diego, CA, USA \quad
  $^{4}$Imperial College London, UK%
}

\begin{document}
\maketitle
\suppressfloats[t]

\begin{abstract}
Convolutional EEG decoders are trained at a fixed width, usually set by their authors on other data. Growing methods add neurons during training where the loss could decrease the most, but
whether they improve compared to a reference width is untested on EEG. Here, we grow three convolutional backbones on 12 motor-imagery datasets under
three protocols and compare each with its reference model per subject. The growing ShallowFBCSPNet scores $2.9$ points above its reference model with only half the parameters ($0.57\times$), SCCNet changes by at most $1.2$ points. Deep4Net growing models show decreased accuracy, but they require adaptation that prevent to compare faithfully the results.

These differences follow the selection step, which keeps a candidate neuron relying on a dynamic threshold from singular values decomposition.
Overall, these results suggest that growth helps when its criterion can rank
the candidate neurons, and that the rate of skipped neuron addition tells where a decoder can be
grown small from scratch.

\end{abstract}

\begin{keywords}
EEG, motor imagery, brain--computer interfaces, neural architecture growth, deep learning
\end{keywords}

\section{Introduction}
\label{sec:intro}

\begin{figure}[t]
\centering
\includegraphics[width=\columnwidth]{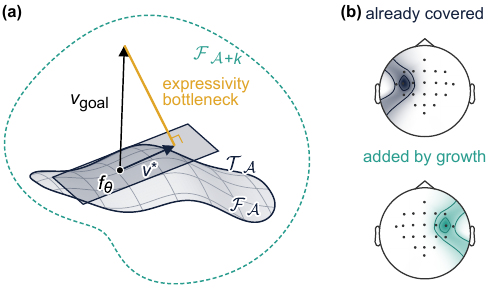}
\vspace{-6pt}  
\caption{(a) The desired update $v_{\mathrm{goal}}$ splits into a reachable part and a
residual, the expressivity bottleneck, which growth covers. (b) On EEG, both read as scalp patterns (schematic).}
\label{fig:bottleneck}
\end{figure}

Convolutional networks are the standard end-to-end decoders for EEG (ShallowFBCSPNet
and Deep4Net~\cite{schirrmeister2017deep}, SCCNet~\cite{wei2019sccnet},
EEGNet~\cite{lawhern2018eegnet}). The large brainwave foundation models proposed since gain about one point over these
small architectures on BCI benchmarks, for a thousand times more
parameters~\cite{lee2025lbm, kuruppu2026review, guetschel2026openeegbench}. The width of these decoders, the number of filters per layer, is tuned by their
authors on one dataset and reused on every other, including in cross-dataset
benchmarks~\cite{jayaram2018moabb, chevallier2024moabb, eder2024benchmarking}. That choice is domain sensitive~\cite{xu2020crossdataset, aristimunha2023transfer}. EEG datasets differ in subjects,
trials and classes, and architecture searches on EEG are tailored to a task or a
subject~\cite{duan2022ctnas, wang2026fbnas}. 

Indeed, one could adapt the width per dataset~\cite{elsken2019nas}, at the cost of multiplying the training cost,
as BCI applications call for decoders that are compact and effective from the
start~\cite{lawhern2018eegnet, khazem2021calibration, wimpff2025finetuning}. Growing methods offer an
alternative approach: the network starts narrow, and training decides where and how much to
widen it~\cite{liu2019splitting, wu2020firefly, errica2025adaptive}.
Within this approach, functional methods~\cite{Verbockhaven2024growing, evci2022gradmax} implement growth steps that reduce the loss most at first order
(Fig.~\ref{fig:bottleneck}a), and a line search selects the number of neurons to add. 
It builds on Net2Net~\cite{chen2016net2net} and extends to directed acyclic graphs~\cite{Douka2025strategies}, with evidence mostly from image benchmarks.

On EEG, it has so far been applied to a classification head~\cite{Velut2026adaptive}
or triggered when learning stalls~\cite{lee2025expansion}. Whether growing the convolutional stages improves on the reference width, and for which
backbones, remains unknown.

To address this question, we grow three convolutional backbones from a narrow start on twelve motor-imagery datasets under three protocols, 34,596 fits in
all, and compare each with its reference model on the same subjects. Our results show
that:
{\setlength{\topsep}{2pt}
\begin{enumerate}
  \setlength{\itemsep}{0pt}
  \setlength{\parskip}{0pt}
\item A decoder grown from scratch can beat its reference width with fewer
  parameters. ShallowFBCSPNet gains 2.9 points within-session with $0.57\times$ the
  parameters and SCCNet stays within 1.2 points of its reference. Differences in architecture and
capacity prevent reaching any strong conclusion regarding Deep4Net and its growing surrogate. 
\item \begin{samepage}Whether growth will help can be read during training.
  Growing ShallowFBCSPNet stabilizes at roughly half the reference model width, the deep backbone reaches the same width as the reference Deep4Net.\end{samepage}
\item Growth thus works best when its criterion can rank the candidate neurons.
  Compared to the final width, the skipped neuron addition rate tells where a decoder can be
  grown small from scratch.
\end{enumerate}}

\section{Methods}
\label{sec:method}

\subsection{The growth step}

We consider a convolutional decoder trained on the trials of one fold. A
\emph{growable layer} is a pair of consecutive convolutions with at most a
normalisation between them. Kernels (channels in PyTorch terminology) can then be added to the output of the first and
to the input of the second without modifying any other layer. The unknown is the width of
considered layer, which starts narrow and is widened during training following the TINY method~\cite{Verbockhaven2024growing}. 
Every five epochs, the layer is offered a growth opportunity, which runs in four steps.

\noindent\textbf{Where to grow.} Let $v_{\mathrm{goal}}$ be the change in the layer's
output that would most decrease the loss at first order, computed per sample. Its
projection $v^*$ onto the tangent space is the best move in the parameter space, and the residual $v_{\mathrm{goal}}-v^*$ is
the expressivity bottleneck of the layer (Fig.~\ref{fig:bottleneck}a).
Second-order moments of the layer's inputs and their cross-covariance with the residual are accumulated over the training set.

\noindent\textbf{Which neuron to grow.} 
Fitting new neurons to the residual is then a low-rank approximation problem: a singular value decomposition returns candidate neurons in decreasing order of
first-order contribution, with singular values $\lambda_1\ge\lambda_2\ge\dots$
We keep the candidates whose singular value is at least
$10\,\%$ of $\lambda_1$. This adapts the method of~\cite{Verbockhaven2024growing}, whose absolute threshold on EEG fell either above a backbone's whole spectrum or below all of it. The relative floor applies the same criterion as a
fraction of the spectrum. A width cap stops the layer at its target width
(Table~\ref{tab:layers}). A growing backbone therefore never ends wider than its
reference model.

\noindent\textbf{How to initialize neurons.} The selected block is added with a scaling factor $s$
chosen by line search over $\{0,\,0.1,\,0.5,\,1.0\}$. Since the candidates were fitted
to the training signal, the search is scored on the last fifth of the epoch's training
batches, withheld from the statistics, using the chosen loss function (cross-entropy in this study). 
When an epoch has fewer than four batches, as
in most within-session fits, it falls back to the training batches.

\noindent\textbf{How many neurons to add.} If the line search returns $s=0$, the update is deleted and
the network is unchanged, which we call a \emph{skipped neuron addition}. Growth continues, with statistics re-estimated at the next opportunity, and a backbone can skip the neuron addition at all 39 opportunities of a 200-epoch fold, ending with its unchanged starting width.

\subsection{Three backbones}
\label{sec:backbones}

\begin{table}[t]
\centering
\caption{Backbones, the layer that grows and its width range. Only the first two rows are width-matched.}
\label{tab:layers}
\vspace{2pt}  
\small  
\setlength{\tabcolsep}{3pt}  
\begin{tabular}{@{}llcl@{}}
\toprule
Backbone & Growable layer & Width & Reference \\
\midrule
Shallow & temporal $\to$ spatial      & $8\!\to\!40$ & ShallowFBCSPNet, 40 \\
SCCNet  & spatial $\to$ spatio-temp.  & $4\!\to\!22$ & SCCNet, 22 \\
Deep    & conv2a $\to$ conv2b & $8\!\to\!32$ & Deep4Net, 4 stages \\
\bottomrule
\end{tabular}
\end{table}
In ShallowFBCSPNet, the temporal and spatial convolutions are chained with no
nonlinearity between them, and the temporal filters grow from 8 towards the reference width of 40.
SCCNet is spatial-first, with only a BatchNorm between its two convolutions, which
grows with the layer. Its spatial components grow from 4 towards the reference width of 22. The
downstream layers (normalisation, readout, classifier) are those of the reference
models. 
They are re-implemented together with the layer, so the comparison is one between
two codebases as well as two procedures.

Deep4Net cannot be grown faithfully, because our growth implementation did not support the intervening pooling operations in the selected deeper blocks, changing the spatial dimension that prevents inclusion of growable layers.
Our deep backbone is therefore a VGG-style~\cite{wang2026wheretogrow} surrogate with two stages, pooling only at the end
of a stage and the layer inside stage two. 
We therefore evaluated a two-stage surrogate against Deep4Net as an exploratory, non-architecture-matched comparison. 
We call the backbone \emph{Deep} rather than Deep4Net.

\subsection{Datasets and protocols}
\label{sec:setup}

\begin{figure*}[t]
\centering
\includegraphics[width=\textwidth]{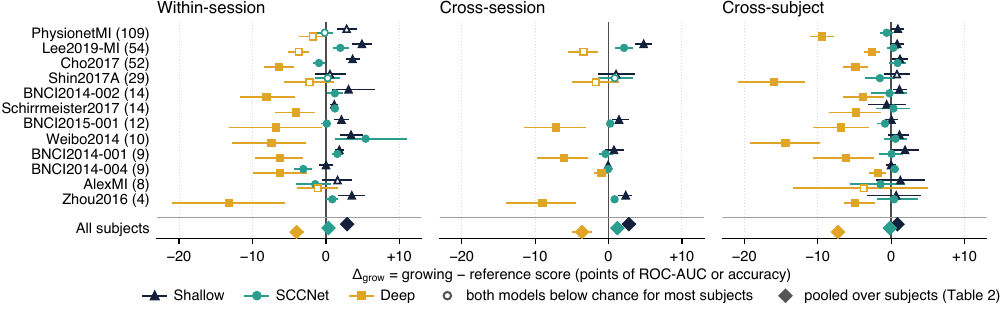}
\caption{
The growing Shallow implementation generally outperformed its reference, whereas the growing Deep surrogate underperformed the architecturally different Deep4Net.
Growing minus reference score, in points, per dataset (rows, subjects in parentheses)
and protocol (panels), with 95\,\% bootstrap intervals over subjects. Bottom row: all
subjects pooled (Table~\ref{tab:contrasts}). Hollow markers: both models below chance for
most subjects.}
\label{fig:radar}
\end{figure*}

We evaluate growth on twelve motor-imagery datasets from MOABB~\cite{jayaram2018moabb}:
AlexMI, BNCI2014-001, BNCI2014-002, BNCI2014-004, BNCI2015-001, Cho2017, Lee2019-MI,
Physio\-net\-MI, Schirr\-meis\-ter2017, Shin2017A, Weibo2014 and Zhou2016. They have two
to four classes and between four and 109 subjects each. The task is to classify the
imagined movement of each trial, band-pass filtered between 8 and 32\,Hz and resampled
to 250\,Hz. Following MOABB, the score is ROC-AUC for the six
two-class datasets and accuracy for the six others, both in percent, so differences between models are in
points.

We evaluate three protocols: within-session (5-fold cross-validation within a
session), cross-session (leave-one-session-out) and cross-subject
(leave-one-subject-out). Each backbone contributes two models, six in all: the
reference model, at the width its authors published, and its growing counterpart, which starts narrow. All models share the same 200-epoch budget and optimiser (AdamW, learning
rate $6.25\times10^{-4}$, batch size 64), with 20\,\% of the training trials held out
for validation, no early stopping and three seeds per fold. Each fit is scored at the
epoch of best validation accuracy.

We report raw scores, without alignment. Euclidean alignment~\cite{junqueira2024ea}, run as a
second complete condition, lowers the scores of all six models by 2.4 to 5.1 points
within-session and raises them by 1 to 2 points only cross-subject. Taking it as the default would have favoured the growing backbones cross-subject and
hidden a loss on every model within-session.

\subsection{Statistical analysis}
\label{sec:stats}

Sessions and seeds are averaged before any test, so the unit of analysis is the
subject, and the $34{,}596$ scored folds reduce to $9{,}168$ subject-level units across
protocols, models and alignment conditions. Treating repeated evaluations of the same subject as independent would overstate precision. For each protocol and backbone, $\dgrow$ is the subject-paired difference
between the growing backbone and the reference model. We report its mean with a 95\,\% percentile
bootstrap interval over subjects ($20{,}000$ resamples) and a Wilcoxon signed-rank
test, Holm-corrected over the nine protocol-by-backbone cells as one family, together
with the minimum detectable effect (MDE) of each cell at 80\,\% power.
\section{Results}
\label{sec:results}

\subsection{The growing backbone is ahead for Shallow and behind for Deep}

To test whether growth improves on the reference width, we compare each growing backbone
with its reference model on every dataset (Fig.~\ref{fig:radar}). The growing Shallow is above zero on all twelve datasets within-session and
on ten of twelve cross-subject, so its gain does not come from a single dataset. The
growing SCCNet changes sign from one dataset to the next. The growing Deep is behind on every dataset in all three protocols.

Table~\ref{tab:contrasts} pools the subjects and gives $\dgrow$ for the nine cells,
next to their MDE. Two cells, both for SCCNet, fall below the MDE with intervals that
cross zero, and we draw no conclusion from them. Of the other seven, four favour the
growing backbone and three the reference model, and all seven survive Holm correction. The
effects are small next to the spread between runs. The median standard deviation
across seeds is 4.4 points, so the largest gain ($+2.9$) is 0.7 times the spread of
a single configuration. It shows up only when averaged over 324 subjects.

\begin{table}[t]
\centering
\caption{Growth is ahead for Shallow and behind for Deep in every protocol. $\dgrow$:
growing minus reference score, in points. Brackets: 95\,\%
bootstrap intervals over subjects. \emph{Ahead}: subjects on which the growing backbone
scores higher. MDE: minimum detectable effect at 80\,\% power. $\dagger$ marks the two cells
whose effect falls below it.}
\label{tab:contrasts}
\vspace{2pt}  
\small  
\setlength{\tabcolsep}{3.5pt}
\begin{tabular}{@{}llr@{\,}lrr@{}}
\toprule
Protocol & Net & \multicolumn{2}{c}{$\dgrow$ [95\,\% CI]} & MDE & ahead \\
\midrule
\multirow{3}{*}{Within-sess.}
 & Shallow & $+2.9$ & $[+2.3,+3.5]$ & 0.9 & 240/324 \\
 & Deep    & $-4.0$ & $[-4.9,-3.1]$ & 1.3 & 107/324 \\
 & SCCNet  & $+0.4^{\dagger}$ & $[-0.2,+0.9]$ & 0.8 & 172/324 \\
\midrule
\multirow{3}{*}{Cross-sess.}
 & Shallow & $+2.8$ & $[+1.9,+3.7]$ & 1.3 & 85/116 \\
 & Deep    & $-3.6$ & $[-5.1,-2.2]$ & 2.0 & 28/116 \\
 & SCCNet  & $+1.2$ & $[+0.4,+2.1]$ & 1.2 & 72/116 \\
\midrule
\multirow{3}{*}{Cross-subj.}
 & Shallow & $+0.9$ & $[+0.4,+1.3]$ & 0.6 & 195/324 \\
 & Deep    & $-7.3$ & $[-8.2,-6.3]$ & 1.4 & 46/324 \\
 & SCCNet  & $-0.2^{\dagger}$ & $[-0.7,+0.3]$ & 0.7 & 168/324 \\
\bottomrule
\end{tabular}
\end{table}

\subsection{The gain comes with fewer parameters}

\begin{figure}[t]
\centering
\includegraphics[width=0.95\columnwidth]{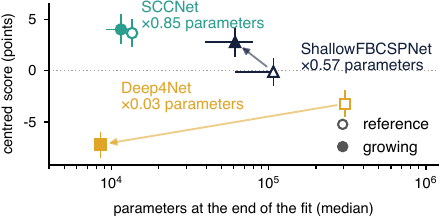}
\caption{The growing Shallow scores higher than its reference with fewer parameters.
Score against size, within-session: scores centred within dataset and averaged over
subjects, against the median parameter count.
Filled markers: growing backbone, open markers: reference model. Horizontal bars:
inter-quartile range of the parameters, vertical bars: 95\,\% bootstrap interval
over subjects.}
\label{fig:pareto}
\end{figure}

To see what the gain costs in size, we place the six models on the accuracy--size
plane (Fig.~\ref{fig:pareto}). The growing Shallow scores higher with fewer parameters. It
reaches $+2.8$ with about $6\times10^{4}$ parameters, against $-0.1$ at
$1.05\times10^{5}$ for the reference model, which is $0.57\times$ the parameters. For SCCNet, the two models nearly coincide ($+4.0$ against $+3.7$).

The network is smaller because growth stops before the target width. Only 29\,\% of
Shallow folds reach width 40, and the median fold is scored at width 26. How far it grows follows the amount of data. By the end of the fit, the share of Shallow fits that reach width 40 rises from none
on Shin2017A, with 16 training trials per fold, to 94\,\% on BNCI2014-001, with 230
(Spearman $\rho=0.77$ over the twelve datasets). Cross-subject, where the training set
pools all other subjects, it reaches 78\,\%. SCCNet reaches its full width of 22 on 78\,\% of folds, so it reproduces the
reference width and its small contrast follows. Growth adds training time, 22\,s per fold for the growing Shallow against 14\,s for
the reference model (medians, same 200-epoch budget).

\subsection{
Skipped neuron addition as a possible accuracy predictor
}
\label{sec:mechanism}
 
What distinguishes the backbone that gains from the ones that do not? The skipped neuron addition
rate does. We define it as the share of growth opportunities at which the line search
returns $s=0$. It orders the three backbones as Table~\ref{tab:contrasts} does. The growing Shallow skips neuron addition at 74.9\,\% of its opportunities, the growing SCCNet at 3.5\,\% and the growing Deep at 0.7\,\%.
Over the $542{,}187$ opportunities of the campaign, Deep adds almost every update it
is offered.

We find the same order in the selection step, which runs before the line search. Of the
neurons proposed at a growth opportunity, Shallow keeps a median of 15\,\% and SCCNet
85\,\%, whereas Deep keeps all of them (100\,\%). The candidate spectra explain this order. They span 3.1 to 4.0 decades
for Shallow and 0.9 to 3.3 for SCCNet, but only 0.3 to 1.4 for Deep. On a spectrum this flat, no threshold separates the candidates and every proposal
goes in.

A single growth step changes little. Its predicted first-order gain is a median of
$10^{-3.6}$ of the gain of one ordinary epoch for Shallow, and $10^{-1.0}$ for Deep.
The held-out loss one epoch before and after a step is centred on zero for
all three backbones. Any effect of growth must therefore come through the training
trajectory it induces~\cite{caillon2024growing}.

A simpler explanation, that a width chosen on a large corpus is too large for one
session, does not hold: binned by training-set size, the growing backbone has no advantage
in the smallest bins and is behind in every bin cross-subject.

Overall, these results suggest that skipped neuron addition, more than the added
capacity, drives the difference between backbones.

\section{Discussion}
\label{sec:discussion}

The present study asked whether growing the convolutional stages of an EEG decoder
improves on its reference width, and for which backbones. It makes two
contributions. First, we compared three growing backbones with their reference
models, subject by subject, on twelve motor-imagery datasets (Fig.~\ref{fig:radar}).
Growth helps ShallowFBCSPNet, which ends ahead with $0.57\times$ the parameters and
changes SCCNet by at most 1.2 points.
The growing deep surrogate underperformed Deep4Net, but differences in architecture and capacity prevent attributing this deficit to growth only.
Second, we showed that these outcomes follow how often the growth procedure
skips neuron addition, a rate that needs no test data. ShallowFBCSPNet skips it at 74.9\,\% of its opportunities, the deep backbone at almost none.

These results complement the evidence for neuron-level growth on image
benchmarks~\cite{Verbockhaven2024growing, evci2022gradmax}, and are consistent with
the growth of a c-VEP classification head~\cite{Velut2026adaptive} and with the pruning of a motor-imagery CNN at nearly equal
accuracy~\cite{vishnupriya2021compressed}. On EEG, the benefit of adapting the width appears as a
smaller network more than as a higher score, which is what efficient BCI decoders
aim for~\cite{wang2024mibminet}. At the other end of the scale,
brainwave foundation models buy about one point with a thousand times more
parameters~\cite{lee2025lbm}. Effects of a few points are usual on this data, where a nine-subject dataset cannot separate two architectures~\cite{kollod2023deep}.

The mechanism we propose is selection. Where the candidate spectrum spans several decades, the first-order estimation can rank the
neurons and rejects most of them, and the network stays small and improves. Where it
is flat, every update is accepted and the network grows to its cap without benefit.
The growth literature asks when and where to add capacity. Layer-growing policies trigger on a fitting risk~\cite{wu2024whentogrow}, and a recent criterion detects
under-expressive layers before growing them~\cite{wang2026wheretogrow}. The skipped neuron addition rate and the span of the candidate spectrum answer both from the
training signal of a single run.

\looseness=-1 
Overall, growth is usually presented as a way to find the right capacity. Here, it
helped where it skipped neuron addition most often. Reported with the final width, the skipped neuron addition
rate would tell where a decoder can be grown small from scratch, and paves the way to
decoders sized by their data, not their authors.

\clearpage
\section{Compliance with Ethical Standards}
This study used human EEG data released in open access by the original studies
and accessed through MOABB~\cite{jayaram2018moabb}. Ethical approval was
not required, as confirmed by the licences of the open access data.

{\small
\let\originalbibliography\thebibliography
\renewcommand{\thebibliography}[1]{\originalbibliography{#1}\setlength{\itemsep}{0pt}}
\bibliographystyle{IEEEbib}
\bibliography{refs}
}

\end{document}